\documentclass[%
reprint,
superscriptaddress,
amsmath,amssymb,
aps,
prapplied,
floatfix,
]{revtex4-2}

\usepackage{amsmath,latexsym}
\usepackage{xcolor}
\usepackage[
  colorlinks=true,
  urlcolor=blue,
  linkcolor=blue,
  citecolor=blue
]{hyperref}

\usepackage[utf8]{inputenc}
\usepackage[english]{babel}
\usepackage{blindtext}
\usepackage{libertine}
\usepackage{libertinust1math}
\usepackage{graphicx}
\usepackage{dcolumn}
\usepackage{bm}
\usepackage{upgreek}
\usepackage{braket}
\usepackage{comment}
\usepackage{ulem}
\usepackage{cleveref}

\begin{document}

\preprint{APS/123-QED}

\title{Temporal filtered quantum sensing with the nitrogen-vacancy center in diamond}

\author{Florian B\"ohm}
\affiliation{Beijing Academy of Quantum Information Sciences, 100193, Beijing, China}
\author{Yan Liu}
\email[Author e-mail address: ]{liuyan@baqis.ac.cn}
\affiliation{Beijing Academy of Quantum Information Sciences, 100193, Beijing, China}
\author{Chengliang Yue}
\affiliation{Beijing Academy of Quantum Information Sciences, 100193, Beijing, China}
\author{Xianqi Dong}
\affiliation{Beijing Academy of Quantum Information Sciences, 100193, Beijing, China}
\author{Huaxue Zhou}
\affiliation{Beijing Academy of Quantum Information Sciences, 100193, Beijing, China}
\author{Dong Wu}
\affiliation{Beijing Academy of Quantum Information Sciences, 100193, Beijing, China}
\author{E Wu}
\affiliation{State Key Laboratory of Precision Spectroscopy, East China Normal University, 200062, Shanghai, China}
\author{Renfu Yang}
\email[Author e-mail address: ]{yangrf@baqis.ac.cn}
\affiliation{Beijing Academy of Quantum Information Sciences, 100193, Beijing, China}

\date{\today}


\begin{abstract}

Nitrogen-vacancy (NV) centers in diamond are among the leading solid-state quantum platforms, offering exceptional spatial resolution and sensitivity for applications such as magnetic field sensing, thermometry, and bioimaging. However, in high background environments—such as those encountered in in vitro diagnostics (IVD), the performance of NV-based sensors can be compromised by strong background fluorescence, particularly from substrates such as nitrocellulose (NC). In this work, we analytically and experimentally investigate the use of pulsed laser excitation combined with time-gating techniques to suppress background fluorescence and enhance the signal-to-noise ratio (SNR) in NV-based quantum sensing, with an emphasis on spin-enhanced biosensing. Through experimental studies using mixed ensembles of silicon-vacancy (SiV) and NV centers in bulk diamond, as well as fluorescent nanodiamonds (FNDs) on NC substrates, we demonstrate significant improvements in NV spin resonance visibility, demonstrated by an increase of the SNR by up to $4\times$, and a resulting measurement time reduction by $16\times$. The presented technique and results here can help significantly increase the readout efficiency and speed in future applications of NV centers in high-background environments, such as in IVD, where the NV centers are used as a fluorescent label for biomolecules.


\end{abstract}

\maketitle

\section{Introduction}

Nitrogen-vacancy (NV) defect centers in diamond \cite{Gruber1997,Rondin2014} have emerged as a powerful platform for quantum sensing due to their unique combination of optical and spin properties. Their exceptional sensitivity to magnetic fields, temperature, and electric fields makes them ideal for a manifold of sensing applications, such as, e.g., magnetic field sensing \cite{Maze2008, Taylor2008, Balasubramanian2008,Rembold2020,Segawa2023}, temperature sensing \cite{Fujiwara2020,Fujiwara2020a}, or enhanced detection in biological labeling \cite{Miller2020, Zhang2021,Li2022b,Oshimi2023,wu2024nanoscale}.

Due to the level system of the NV center and its spin-dependent branching ratio, the current spin state of the NV center's electron spin can be optically read out \cite{jelezko2004observation}. 
Although most of the groundbreaking research using NV centers is done in controlled, low-background environments, such as high-quality bulk CVD diamond, this approach faces challenges in sensing or detection scenarios, where the environment of the NV center can not be controlled to exhibit low background fluorescence. 
Such scenario can, for example, occur in biological environments \cite{Berezin2010}, or the recent significant progress of using fluorescent nanodiamonds for ultrasensitive in vitro diagnostics (IVD) \cite{Miller2020,Li2022b}. Thus, a general understanding of the influence of the intensity and lifetime of background fluorescence on the NV spin state readout, and the development of mitigation strategies for noisy environment is crucial for future applications of NV sensing.

As it is well known, the  excited state lifetime of the NV center shows a spin-dependent behaviour \cite{Batalov2008,Neumann2009}, and based on that, recent efforts have been made to develop new spin-state detection protocols, such as to enhance the spin-state readout robustness by using the fitted excited-state lifetimes \cite{sturner2021magnetometry,cao2025detection}. Furthermore, previous studies have already explored methods using pulsed laser excitation and time-resolved photon-counting for enhancing the detection or imaging of NV centers from high background noise conditions such as in cell and tissue \cite{Chen2019,Faklaris2008a,Hui2014wide}. However, there remains a gap in comprehensive studies addressing the optimization of time-filtering techniques to enhance spin discrimination signal-to-noise ratio (SNR) and magnetic field sensitivity, particularly in conjunction with varying properties of the background fluorescence properties, such as lifetime and intensity, and also in combination with varying the excitation laser repetition rate for optimal readout contrast. Therefore, in this work, we show a detailed description of how pulsed excitation and time-correlated single-photon counting (TCSPC) can significantly improve SNR and magnetic field sensitivity by effectively filtering out fast-decaying background light, as well as demonstrate a simple hardware integration without the need for fast detection hardware.

In the following, we first illustrate the principle of time-filtering for enhancing solid-state spin readout sensitivity in high-background environments using pulsed excitation and temporal filtering, and show the experimental implementation in bulk diamond, which exhibits a strong fast decay component due to silicon-vacancy (SiV) centers and a slower decay component due to NV centers. 
Additionally, we show a practical example of how time-gating can further enhance the detection of fluorescent nanodiamonds on a nitrocellulose (NC) strip for in vitro detection. Here, we additionally demonstrate the hardware implementation of time-gating, avoiding the requirement of TCSPC and data post-procession. 
We can demonstrate that in both cases a significant improvement in spin-readout SNR and magnetic field sensitivity can be achieved by time-gating of the detected fluorescence, combined with optimizing the pulsed laser repetition rate, and how time-gating can significantly reduce the required detection time. 
Furthermore, unlike methods such as MW power broadening \cite{cao2025detection}, temporal filtering increases ODMR contrast without affecting line width $\Delta\nu$ (\autoref{fig:fig2}(a)), preserving spectral resolution.


\section{Theory of time-gating for NV spin readout}

\subsection{NV Spin-Dependent Fluorescence}

\begin{figure}[htbp]
  \centering
  \includegraphics[width=1\columnwidth]{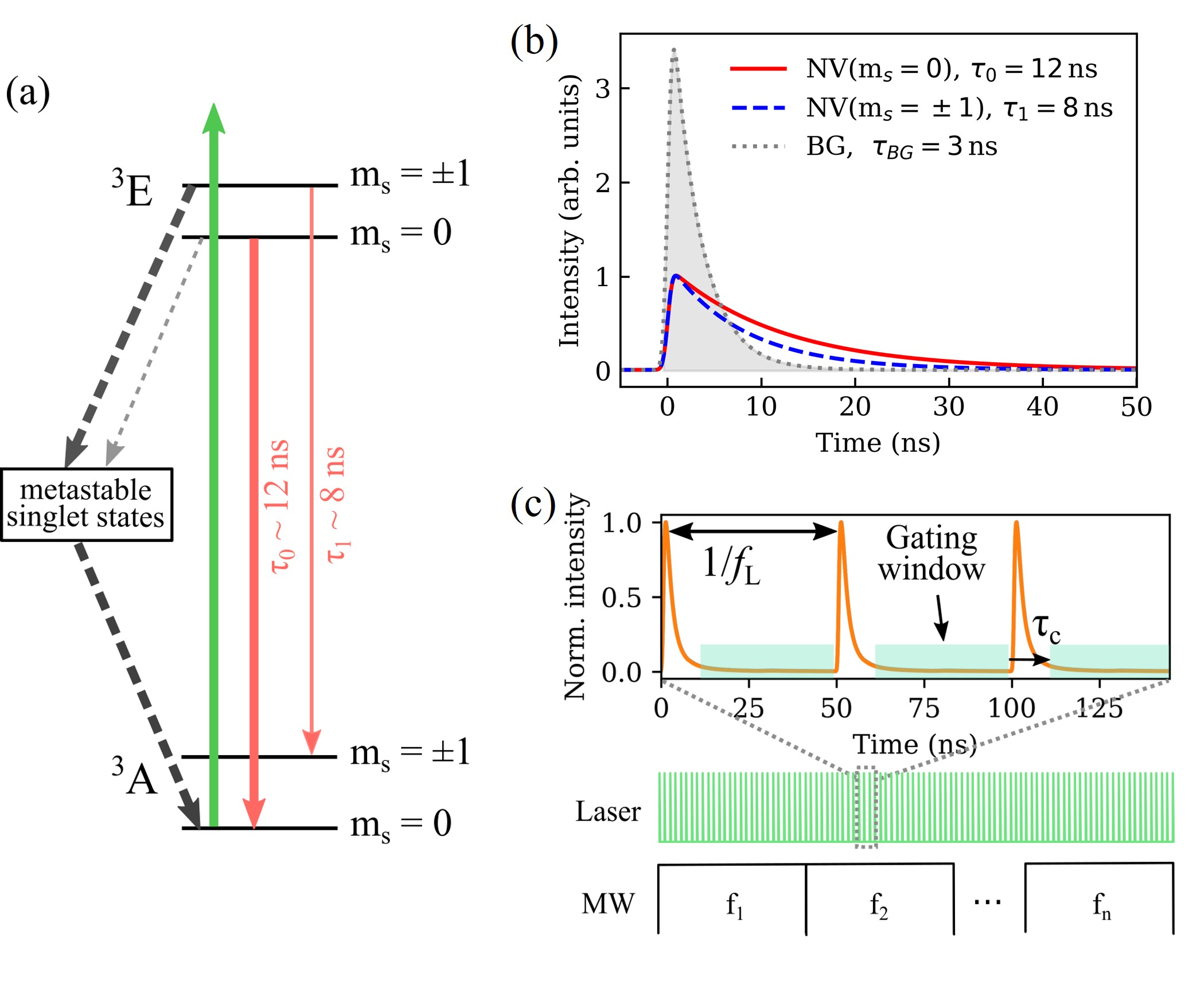}
\caption{
(a) Simplified level structure of the NV center, showing the effective lifetime $\tau$ of the spin sublevels $m_S = 0$ and $m_S = \pm 1$, due to the excited state decay rate being the sum of the radiative decay rate and inter-system crossing rate (solid lines show radiative transitions, and broken lines non-radiative transitions). 
(b) Analytic representation of decay curves with different fluorescence lifetimes $\tau_{m_S = 0}=12\,$ns (red solid line), and $\tau_{m_S = \pm1}=8\,$ns (blue dashed line) of the $m_S=0$ and $m_S=\pm1$ spin sublevels, as well as the fast decay curve of a background source with a shorter lifetime of $\tau_\text{BG}=3\,$ns (grey dotted line and the shaded area). The temporal differences allow for high contrast detection of spin resonance with proper time-gating.
(c) Principle of pulsed laser CW-ODMR and temporal filtering. The detected TCSPC curve shows the experimentally detected signal from a sample which exhibits NV fluorescence, as well as a strong short-lifetime background. 
}
\label{fig:fig1}
\end{figure}

The essential electronic structure of the negatively charged (NV$^-$) center (which from here will be referred to as NV center) at zero external magnetic field  is shown in \hyperref[fig:fig1]{\autoref{fig:fig1}(a)}. The electronic spin ground and excited states form triplet manifolds, with a zero-field energy splitting between the ground state (GS) $m_S=0$ and $m_S=\pm1$ spin sublevels of $D_\text{GS}=2.87\,$GHz, with the $m_S=\pm1$ spin sublevels degenerate at zero external magnetic field. 
The main radiative and non-radiative transitions that can occur between the excited ($^3E$) and ground ($^3A$) states, are denoted by the solid and dashed arrows, respectively. 
The solid green arrow indicates the off-resonant excitation by a green laser, and the red arrows indicate the resulting broad red fluorescence at around $(600-800)\,$nm. 
Although the $m_S=0$ and $m_S=\pm1$ states have identical oscillator strength \cite{Robledo2011}, the non-radiative transitions to metastable singlet states from the excited levels are stronger from the $m_S=\pm1$ states than from the $m_S=0$ state (indicated by the strength of the dashed arrows in \hyperref[fig:fig1]{\autoref{fig:fig1}(a)}). 
Due to this spin-dependent branching ratio, the NV center's fluorescence intensity is spin-dependent. When continuously cycling the spin transitions, the resulting fluorescence intensity from the $m_S=\pm1$ states is typically up to $15\%$ lower than from the $m_S=0$ state \cite{Barry2020}.
Since the excited-state decay rate is the sum of the radiative decay rate and the intersystem crossing rate, the effective lifetime of the $m_S=\pm1$ states is significantly shorter than that of the $m_S=0$ state; however, the amplitudes of their respective excited-state lifetime decay curves remain equal \cite{sturner2021magnetometry,cao2025detection}.

In bulk diamond, typically lifetimes of the $\tau_{m_S = 0}\approx12\,$ns, and $\tau_{m_S = \pm1}\approx8\,$ns are reported \cite{Batalov2008,Doherty2013,sturner2021magnetometry}, while even longer lifetimes are reported in diamond nanocrystals (e.g., $\tau_{m_S = 0}\approx23\,$ns, and $\tau_{m_S = \pm1}\approx13\,$ns \cite{Neumann2009}).  
The expected resulting TCSPC decay curves from a single NV center's $m_S=0$ and $m_S=\pm1$ spin projection in bulk diamond are shown in \hyperref[fig:fig1]{\autoref{fig:fig1}}(b), (monoexponential decay for the NV lifetime, convoluted with a Gaussian distribution of width $\sigma=0.4\,$ns representing the typical instrument response function).

\subsection{Influence of Background Fluorescence}

The spin-dependent fluorescence of NV centers enables optical detection of electron spin resonance via optically detected magnetic resonance (ODMR) \cite{Doherty2013}. In continuous-wave ODMR (CW-ODMR), the NV center is continuously excited with an off-resonant laser and simultaneously driven with microwaves (MW), establishing a balance between laser-induced spin polarization ($m_S=0$) and MW-driven transitions to $m_S=\pm1$ \cite{Barry2020,hall2016detection,wang2022picotesla}. While pulsed ODMR methods offer high sensitivity and resolution, CW-ODMR remains favorable for robust, ambient-compatible applications such as temperature sensing \cite{Fujiwara2020,Fujiwara2020a} or ensemble-based detection with fluorescent nanodiamonds (FNDs) \cite{Miller2020,Li2022b,Oshimi2023}.

In many environments, background fluorescence, e.g., from biological tissue, substrates, or other diamond defects can overlap with NV fluorescence and reduce sensitivity. However, as the NV center’s excited-state lifetime is spin-dependent and relatively long ($\sim10\,$ns), compared to the typically short (few ns) lifetime of background signals \cite{Berezin2010,Shah2017,Bradac2019}, this distinction allows for temporal filtering. In short, photons detected in the initial few nanoseconds after laser excitation are discarded, suppressing short-lived background while retaining the longer-lived spin-dependent NV fluorescence \cite{Faklaris2008a,Hui2014wide,Batalov2008}.

\subsection{Temporal Filtering}
In the following we will introduce the mathematical concept of time-gating, and how it can be used to improve the spin-state readout of NV centers. The photon counts from a single excitation cycle can be expressed as:

\begin{equation}
\label{eq:photon_counts_ni_perpulse}
n_{i} = \int_{t_0}^{t_1} A \cdot e^{-\frac{t}{\tau_i}} \, dt = A \cdot \tau_i \cdot (e^{-\frac{t_0}{\tau_i}}-e^{-\frac{t_1}{\tau_i}}),
\end{equation}{}
where $A$ is the amplitude of the fluorescence decay curve, and $t_0$ and $t_1$ are the start and end time of the gating window, respectively. The integrated photon counts from the NV center with $m_S=0$ and $m_S=\pm1$ spin states are denoted as $n_{0}$ and $n_{1}$, the background fluorescence is denoted as $n_\text{BG}$. 
The total detected counts for each spin state are then:
\begin{equation}
N_i = n_i + n_{\text{BG}}, \quad i \in {0,1},
\label{eq:Ni}
\end{equation}
and the ODMR readout contrast is defined as:
\begin{equation}
C = \frac{N_0 - N_1}{N_0}.
\label{eq:contrast}
\end{equation}

Temporal gating can suppress short-lived background and enhances spin-state distinguishability, as illustrated in \autoref{fig:fig1}(b–c). 
However, ODMR contrast is not a good indicator for readout improvement, as it can in theory be increased to very large values, while sacrificing valuable photon counts. Therefore, to properly quantify the performance of time-gating, we use the signal-to-noise ratio (SNR):
\begin{equation}
\text{SNR} = \frac{N_0 - N_1}{\sqrt{N_0 + N_1}},
\label{eq:SNR}
\end{equation}
and define the ratio between ungated SNR, and SNR with time-gating applied \textit{SNR enhancement factor (EF$_\text{SNR}$)}. Assuming a perfect separation of the background signal from the NV signal, and $n_1 =(1-C)\cdot n_0$ (with typically $n_1 \gtrsim  0.85\cdot n_0$), the theoretical maximum achievable enhancement of the SNR by filtering out background fluorescence is given by
\begin{equation}
\label{eq:SNR_enhancement}
    EF_\text{SNR}=\frac{\text{SNR}_\text{gated}}{\text{SNR}_\text{ungated}} = 
    \sqrt{1+\frac{2}{(2-C)}\cdot\frac{n_\text{BG}}{n_{0}}}.
\end{equation}

From \autoref{eq:SNR_enhancement} we can see that the SNR increases with the square root of the integration time $\sqrt{t}$, with a fixed laser pulse repetition rate $f_{\text{L}}$.
Hence, the measurement \textit{speedup factor (SF)} shows a quadratic relation to the SNR enhancement factor \cite{Hopper2018}:
\begin{equation}
\label{eq:SNRspeedup}
SF = (EF_\text{SNR})^2=\frac{T_\text{ungated}}{T_\text{gated}},
\end{equation}{}
where $T_\text{ungated}$ is the integration time required to achieve a certain SNR threshold for the original measurement without time-gating, and $T_\text{gated}$ is the integration time required to reach the same SNR when the signal is time-gated.

Finally it should be mentioned, that another important metric in NV ODMR is the magnetic field sensitivity of CW-ODMR, which is given by \cite{Dreau2011,Barry2020}:
\begin{equation}
\eta_{\text{cw}} = \frac{4}{3\sqrt{3}} \cdot \frac{h}{g_e \mu_B} \cdot \Delta\nu \cdot \frac{\sqrt{R_0}}{R_0 - R_1},
\label{eq:sens}
\end{equation}
where $R_0 = r_0 + r_{\text{BG}}$ and $R_1 = r_1 + r_{\text{BG}}$, and $\Delta\nu$ is the resonance linewidth from Lorentzian fits. As it can be seen, this expression is inversly related to the definition of SNR.

\begin{figure}
\centering
  \includegraphics[width=1\columnwidth]{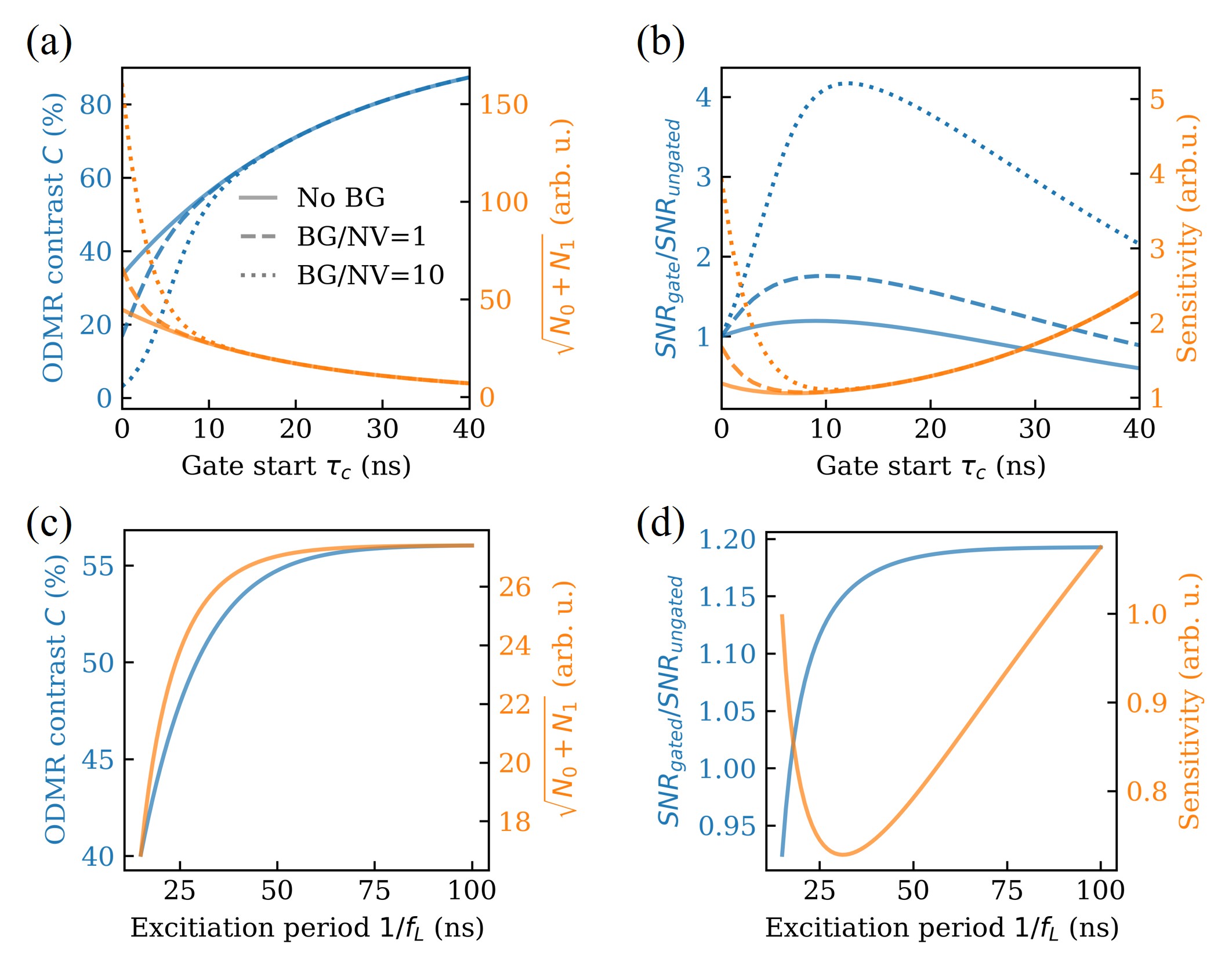}
\caption{Calculated ODMR contrast and sensitivity with different gating window and laser repetition rate, with fluorescence lifetimes $\tau_{m_S = 0}=12\,$ns, and $\tau_{m_S = \pm1}=8\,$ns of the $m_S=0$ and $m_S=\pm1$ spin sublevels, and the fast decay curve of a background source with a shorter lifetime of $\tau_\text{BG}=1.7\,$ns. (a) ODMR contrast and the shot noise component $\sqrt{N_{0} + N_{1}}$ when varying the onset time of the gating window $\tau_\text{c}$, and (b) SNR enhancement calculated with \autoref{eq:SNR} and \autoref{eq:SNR_enhancement}, and the magnetic field sensitivity calculated with \autoref{eq:sens} when varying $\tau_\text{c}$. The noise levels are depicted as the amplitude ratios between the fast decay noise and NV fluorescence decay. Simulated ODMR contrast and shot noise component (c), and the SNR enhancement as well as the magnetic field sensitivity (d) as a function of the laser repetition rate $f_{\text{L}}$ with $\tau_\text{c}=10\,$ns.}
\label{fig:simulations}
\end{figure}

Using the above analytic expressions we show the effects of time-gating in \autoref{fig:simulations}(a-b) for the example presented in \autoref{fig:fig1}, where we assume a NV center with $\tau_{m_S = 0}=12\,$ns, and $\tau_{m_S = \pm1}=8\,$ns, and a fast decaying background source with a shorter lifetime of $\tau_\text{BG}=1.7\,$ns. As discussed, in \autoref{fig:simulations}(a) we can see that the ODMR contrast can be increased, but the effective count-rate is reduced for longer $\tau_c$. Therefore, in
\autoref{fig:simulations}(b) we show the resulting \textit{SNR enhancement factor (EF$_{SNR}$)}, and magnetic field sensitivity $\eta$, from the optimal gate window ($\tau_c$) can be deduced for maximizing SNR and sensitivity.

When applying pulsed excitation, one further needs to consider that the repetition rate is an important parameter, as faster repetiton rates will increase photon count rate, but too fast excitation rates will mainly cycle the fast-decaying background, rendering time-gating ineffective.
The photon rate with gating can be expressed as:
\begin{equation}
r_i = f_{\text{L}} \cdot \int_{\tau_c}^{1/f_{\text{L}}} A e^{-\frac{t}{\tau_i}} , dt,
\label{eq:rate}
\end{equation}
with laser repetition rate $f_{\text{L}}$ and gating onset $\tau_c$.

In \hyperref[fig:fig:simulations]{\autoref{fig:simulations}(c-d)}, we show how the laser repetition rate $f_{\text{L}}$ theoretically affects the ODMR contrast and the shot noise component, as well as the SNR and magnetic field sensitivity.
Increasing the laser repetition rate increases photon yield, but higher rates ($1/f_{\text{L}} \lesssim \tau_0$) may cause saturation effects and reduce gating effectiveness. Thus, for the simplified model of two exponential decays (as introduced before), an optimal laser pulse period of $1/f_{\text{L}} \approx 50\,$ns can be deduced. 
We can however see, that choosing the optimal gating window $\tau_c$ has a much stronger impact on the SNR and sensitivity than the laser repetition rate $f_{\text{L}}$, therefore we will mainly focus optimizing the gating window $\tau_c$ in the following section, where we  experimentally demonstrate optimization of SNR and magnetic sensitivity using the presented gating strategy .

\section{Experimental Implementation}
\subsection{Model System: NV/SiV Ensemble in Bulk Diamond}

The experiments described in the following were carried out by recording the time-trace of the spin-dependent fluorescence of samples with NV centers at room temperature using a home-built confocal microscope. The excitation light source is a pulsed picosecond (pulse length $<100\,$ps) green $532\,$nm laser with a variable repetition rate up to a maximum of $80\,$MHz (\textit{PicoQuant}, LDH-D-FA-530L). The excitation and detection is through the same microscope objective (\textit{Olympus}, MPlanFL N 100x/0.90). The detected fluorescence is separated from the excitation laser by a dichroic mirror (\textit{Chroma}, ZT561rdc), spatially filtered by a $d=40\,$\textmu m pinhole, and additionally spectrally filtered by a band pass filter $(625-792)\,$nm (\textit{Semrock}, FF01-709/167-25) in front of the single photon detection module (\textit{Excelitas}, SPCM-AQRH-14). The arrival time of the photons is recorded by a multiple-event time digitizer (\textit{FastComTec}, MCS8A), with the time resolution set to $100\,$ps. 
The short excitation pulses together with the high time resolution of the detected photons allows  to precisely analyze the fluorescence response received from the sample, in order to perform the lifetime filtering analysis.

To drive transitions between the NV electron spin ground state sublevels, a Vector Signal Generator  (\textit{R\&S}, SMIQ06L) is used to generate the microwave driving field, which was then amplified (\textit{Mini-Circuits}, ZHL-15W-422-S+) and sent to a thin ($\approx50\,$µm) copper wire, soldered upon the investigated diamond sample.

The bulk sample utilized is a (111) oriented diamond film with a thin layer (thickness $(100\sim200)\,$nm) of ensemble NV and SiV centers grown on the top surface. Despite the existence of SiV centers, the grown NV centers were preferentially oriented along [111] direction \cite{Michl2014,Ishiwata2017}. The nanodiamond samples were prepared from milling type Ib high temperature high pressure bulk diamonds, and showed a typical diameter of $200\,$nm. NV centers in the FNDs were created by electron beam irradiation with kinetic energy of $12\,$MeV and subsequent $10\,$hours of annealing under $800^\circ$C in vacuum. 




From \autoref{eq:SNR} it can be seen, that in order to achieve maximal SNR it is important to maximize the difference signal detected from the different spin states, while simultaneously keeping the detection of spin-state unrelated photons low. 
Especially in high-background environments, the SNR can be strongly decreased. As previously discussed, especially for short-lived background noise, this effect can partly be circumvented by pulsed excitation and time-gating of the detected photons in order to separate the background noise from the NV signal.

\begin{figure}[htbp]
  \centering
  \includegraphics[width=1\columnwidth]{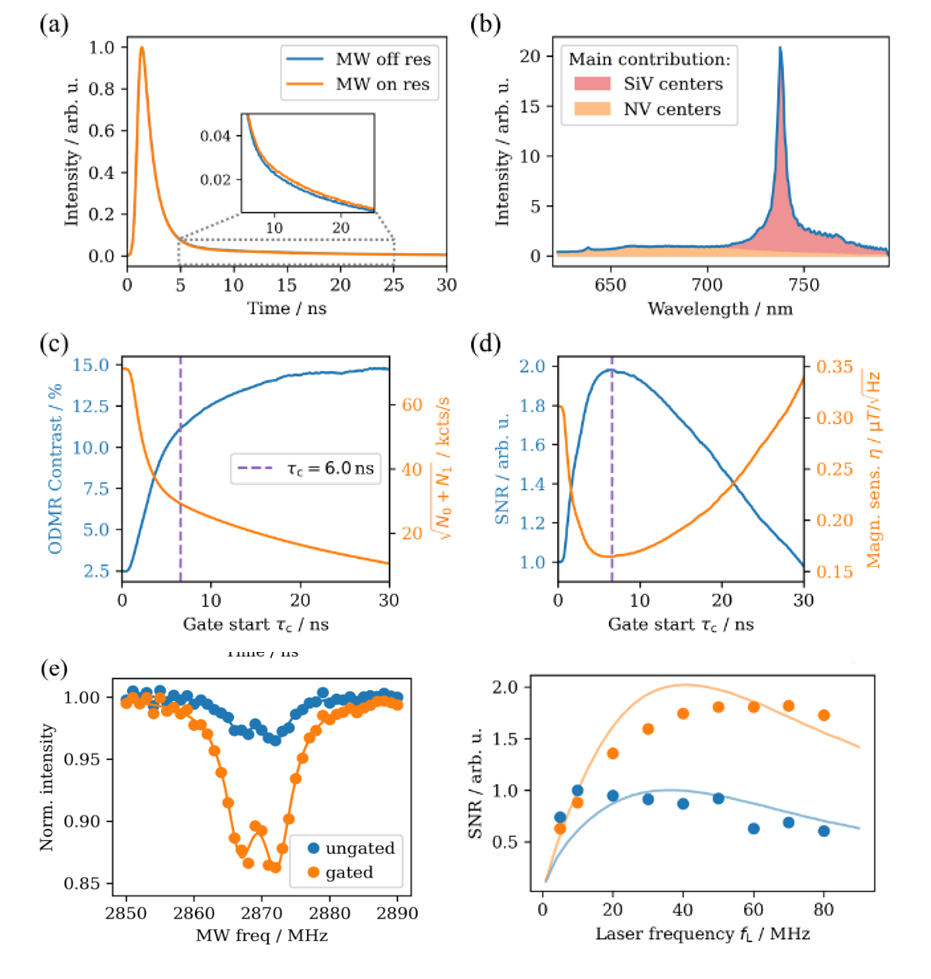}
\caption{
(a)  TSPC signal with contribution from SiV and NV centers. The strong short lifetime component is mainly due to SiV centers ($\tau_\text{SiV}\approx1.7\,$ns), and the long-lifetime component from NV centers ($\tau_\text{NV}\approx(11-12)\,$ns). 
(b) Fluorescence recorded from the sample, showing the contribution of both NV and SiV color centers. 
(c) Resulting ODMR contrast $C = (N_{0} - N_{1})/N_{0}$, and shot noise component $\sqrt{N_{0} + N_{1}}$ when varying the onset time of the gating window $\tau_\text{c}$. The dashed line shows the ideal value of $\tau_\text{c} = 6\,$ns. 
(d) The resulting SNR enhancement $EF_\text{SNR}=\frac{\text{SNR}_\text{gated}}{\text{SNR}_\text{ungated}}$ and magnetic field sensitivity (calculated from \autoref{eq:SNR} and \autoref{eq:sens}, respectively) when varying $\tau_\text{c}$, which both show an enhancement of $\approx 2$ when the time gate $\tau_\text{c}$ is chosen optimally.
(e) CW-ODMR spectra resulting from the ungated and gated detection (data points) shown in (d), as well as double Lorentzian fits to the data (solid lines), showing the strong increase of ODMR contrast ($C_\text{ungated} \lesssim 2.5\% \rightarrow C_\text{gated} \sim 15\%$).
(f) SNR without and with time gating for different laser repetition rates (datapoints), and simplified model (solid lines). More details can be found in \autoref{app:rep_rate}.
}
\label{fig:fig2}
\end{figure}

In \hyperref[fig:fig2]{\autoref{fig:fig2}(a)}, we show the typical TSPC signal observed from the bulk diamond sample with mixed SiV and NV centers, where a strong short lifetime component can be seen, which is mainly due to the SiV color centers (typical lifetime $\tau_\text{SiV} \approx 1.7\,$ns \cite{Sipahigil2014,Zuber2023}), as well as a long lifetime component resulting from NV center fluorescence. The TSPC signal was integrated for approximately $10\,$s, with the MW frequency being switched \textit{on} and \textit{off} for half of the integration time each (the MW is switched with $50\,$Hz in order to avoid any slow drifts or other effects caused by the application of the MW to the sample), and the photon count rate of the experiment is about $R\approx5\cdot10^6\,$cts/s.

From the strong peak around $740\,$nm in the fluorescence spectrum (see \hyperref[fig:fig2]{\autoref{fig:fig2}(b)}) we can confirm that a significant part of the fluorescence stems from SiV centers \cite{Sipahigil2014,Zuber2023}, and integrating the separate parts of the emission spectrum, we can find that the ratio of NV:SiV emission is about 1:3, which applied to \autoref{eq:SNR_enhancement} leads to a possible SNR enhancement of $EF_\text{SNR} \lesssim 2.1$.

In \hyperref[fig:fig2]{\autoref{fig:fig2}(c)}, the resulting ODMR contrast $C = (N_{0} - N_{1})/N_{0}$, and shot noise component $\sqrt{N_{0} + N_{1}}$ when varying the onset time of the gating window $\tau_\text{c}$ is shown. 
With increasing gating window length the contrast can be increased drastically, however, as previously introduced the important metrics are the SNR and magnetic field sensitivity, which can be calculated from the integrated photon counts using \autoref{eq:SNR}. 

The resulting SNR and sensitivity when varying the time-gating window $\tau_\text{c}$ is shown in \hyperref[fig:fig2]{\autoref{fig:fig2}(d)}, where the SNR peaks, and the sensitivity is lowest around $\tau_\text{c} \approx 6\,$ns. At this time-gating window, the SNR shows a maximum enhancement factor of $EF_{\text{SNR}} \approx 2$, leading to a measurement speedup of $SF_{\text{SNR}} \approx 4$ compared to the ungated measurement (see \autoref{eq:SNRspeedup}). 
Likewise, the magnetic field sensitivity shows a decrease by factor $\frac{\eta_\text{ungated}}{\eta_\text{gated}} \approx 2$, showing a similar speedup for magnetic field measurements using time gating in this example.

In \hyperref[fig:fig2]{\autoref{fig:fig2}(e)}, we show the comparison of original ODMR spectrum and the visibility-enhanced ODMR spectrum using the optimal temporal gating parameters ($\tau_\text{c} = 6\,$ns), which shows the predicted increase of ODMR contrast from $C_\text{ungated} \lesssim 2.5\% \rightarrow C_\text{gated} \gtrsim 10\%$.

Lastly, the data points in \hyperref[fig:fig2]{\autoref{fig:fig2}(f)} show the SNR without and with time gating for different laser repetition rates. The excitation laser power was set to be below saturation of the NV center. Furthermore, the time gates have been optimized at each repetition frequency, as discussed before.
The solid lines in \hyperref[fig:fig2]{\autoref{fig:fig2}(f)} show the predicted results from a simplified model, which should merely represent the trend of the experimental data (a more detailed description of the influence of the repetition rate is given in \autoref{app:rep_rate}).

\subsection{Application Example: IVD with FNDs}
In the previous section we could demonstrate  how time-gating can be used to enhance the SNR and magnetic field sensitivity of NV centers in bulk diamond using a model system of NV and SiV ensembles. 
In the following, we will show how this technique can be applied to enhance the detection of NV centers in FNDs on high-background nitrocellulose (NC) substrates, which are e.g. used for in vitro diagnostics (IVD) \cite{Miller2020,Li2022b,Oshimi2023}.

\begin{figure}[htbp]
  \centering
  \includegraphics[width=1\columnwidth]{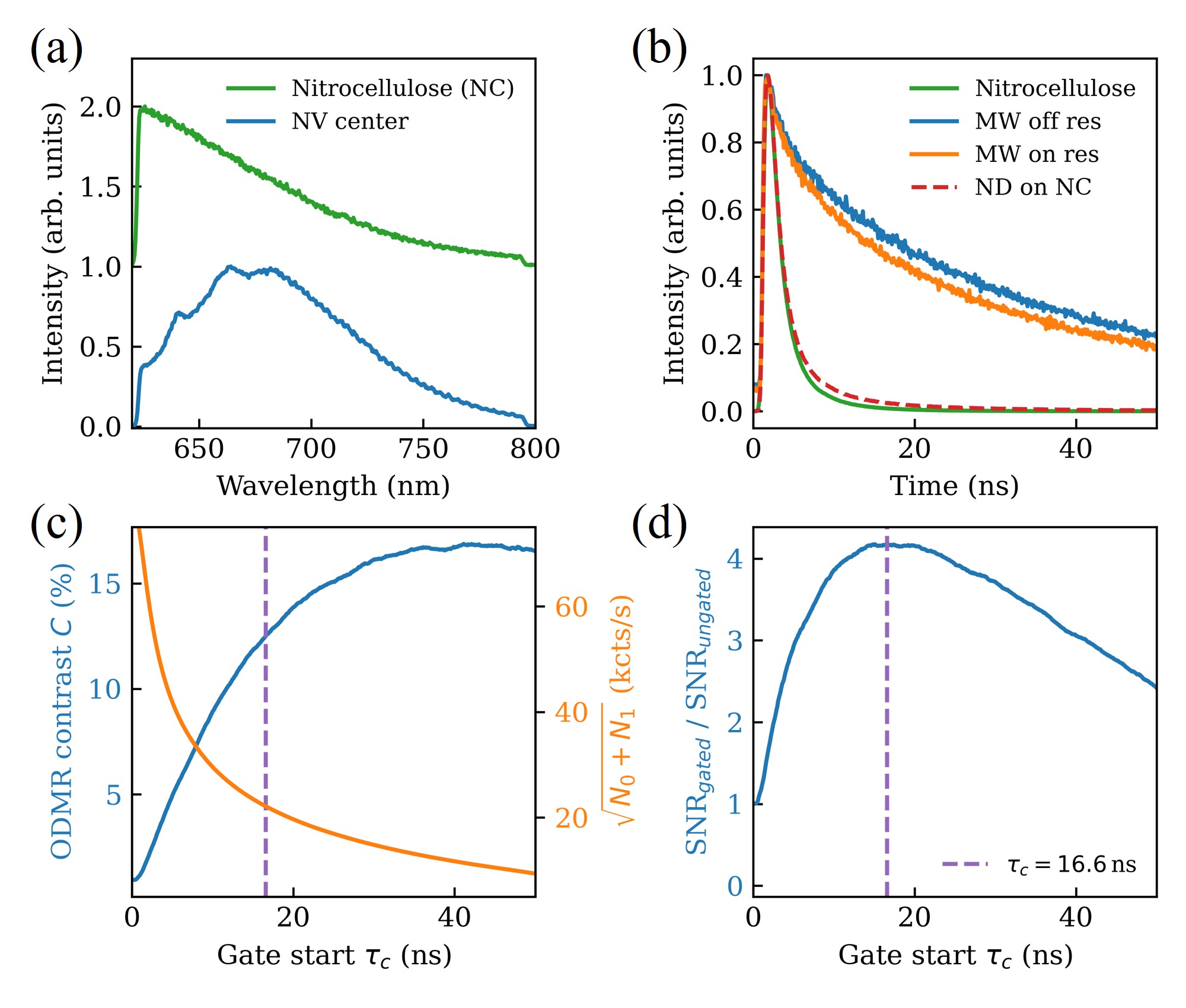}
\caption{(a) Fluorescence spectra of a pristine NC strip (offset by 1), and from a FND, showing mainly NV fluorescence. 
(b) TCSPC curves from the pristine NC strip, the FND measured on glass substrate with the MW either on or off resonance, and FND on the NC strip with MW off resonance. The NC shows a lifetime of about $\tau_\text{NC}\approx4\,$ns and the FNDs show fluorescence lifetimes of about $\tau_\text{FND}\approx(20-30)\,$ns. 
(c) ODMR contrast and shot noise component when varying the onset time of the gating window $\tau_\text{c}$. The dashed line shows the ideal value of $\tau_\text{c} = 16.6\,$ns. 
(d) SNR (calculated from \autoref{eq:SNR}) when varying $\tau_\text{c}$, which shows an enhancement of up to $\approx 4$ when the time gate is chosen optimally.
}
\label{fig:fig4}
\end{figure}

In spin-enhanced IVD, the concept is to increase sensitivity by applying resonant MW frequency to the NV center to confirm the existence of NV centers by observing the spin-related fluorescence change of the NV centers \cite{Miller2020}. 
One challenge in this spin-enhanced IVD is unintentional background fluorescence exhibited by the substrate, which is typically nitrocellulose (NC) \cite{Shah2017}. 
The fluorescence properties of NC have already been reported to show a broad fluorescence spectrum,  overlapping with the emission spectrum of the NV center \cite{Nguyen2020}, which we could confirm by taking fluorescence spectra from pure NC when excited by $532\,$nm as shown in \hyperref[fig:fig4]{\autoref{fig:fig4}(a)}. Due to this overlap of the fluorescence spectra, it is not possible to use spectral filtering in order to efficiently remove the NC background signal.

However, similar to the previously observed SiV background, NC is reported to also exhibit relatively short fluorescence lifetimes of $(2-5)\,$ns \cite{Shah2017}. 
In \hyperref[fig:fig4]{\autoref{fig:fig4}(b)} the normalized TCSPC signal from a blank IVD NC strip, as well as the time-trace observed from a $\approx200\,$nm diameter FND (see details of preparation and characterization in \hyperref[sec:appendix_ND]{Appendix \ref{sec:appendix_ND}}) on a glass substrate can be found. 
The NC fluorescence shows a fast decay with a time constant of around $\tau_\text{NC}\approx4\,$ns, whilst the measurement on the FND shows fluorescence lifetimes of about $\tau_\text{FND}\approx(20-30)\,$ns.
After separately measuring the NC and FND signal, we apply FNDs to the NC strip and record the resulting TCSPC signal (red dashed line in \hyperref[fig:fig4]{\autoref{fig:fig4}}(b)). In this case, the NC has very high contribution to the overall fluorescence intensity, hence it is very challenging to even notice the existence of the NV fluorescence. As a result, the conventional ODMR shows very low contrast ($C \approx 1.2\,\%$).

Considering the distinct excited-state lifetimes of NC and NV, we applied the temporal filtering technique in order to improve the visibility of NV fluorescence signal as well as the ODMR contrast. Just as before with the SiV background, the gating window $\tau_\text{c}$ for achieving optimal SNR can be determined by \autoref{eq:SNR}. In \hyperref[fig:fig4]{\autoref{fig:fig4}(c)}, the effect of the length of the gating window on the ODMR contrast and the shot noise is shown. The resulted spin-related SNR is shown in \hyperref[fig:fig4]{\autoref{fig:fig4}(e)}. Due to the longer lifetime of NC compared to SiV and also the longer lifetime of NV centers in FNDs, the optimal gating window can be determined at $\tau_\text{c} \approx 16.6\,$ns, which is more than twice the length than that in the previous bulk diamond example. This strongly showcases how the optimal gating window is dependent on the properties of the background light. 
In conclusion, although this practical example shows a very strong background noise, the gating protocol can enhance the SNR by up to $EF_{\text{SNR}} \approx 4$, which  leads to a potential measurement speedup of up to $SF_{\text{SNR}} \approx 16$ compared to the ungated measurement (see \autoref{eq:SNRspeedup}).

\subsection{Hardware Time-Gating for Real-Time SNR Enhancement}
\begin{figure}[htbp]
  \centering
  \includegraphics[width=1\columnwidth]{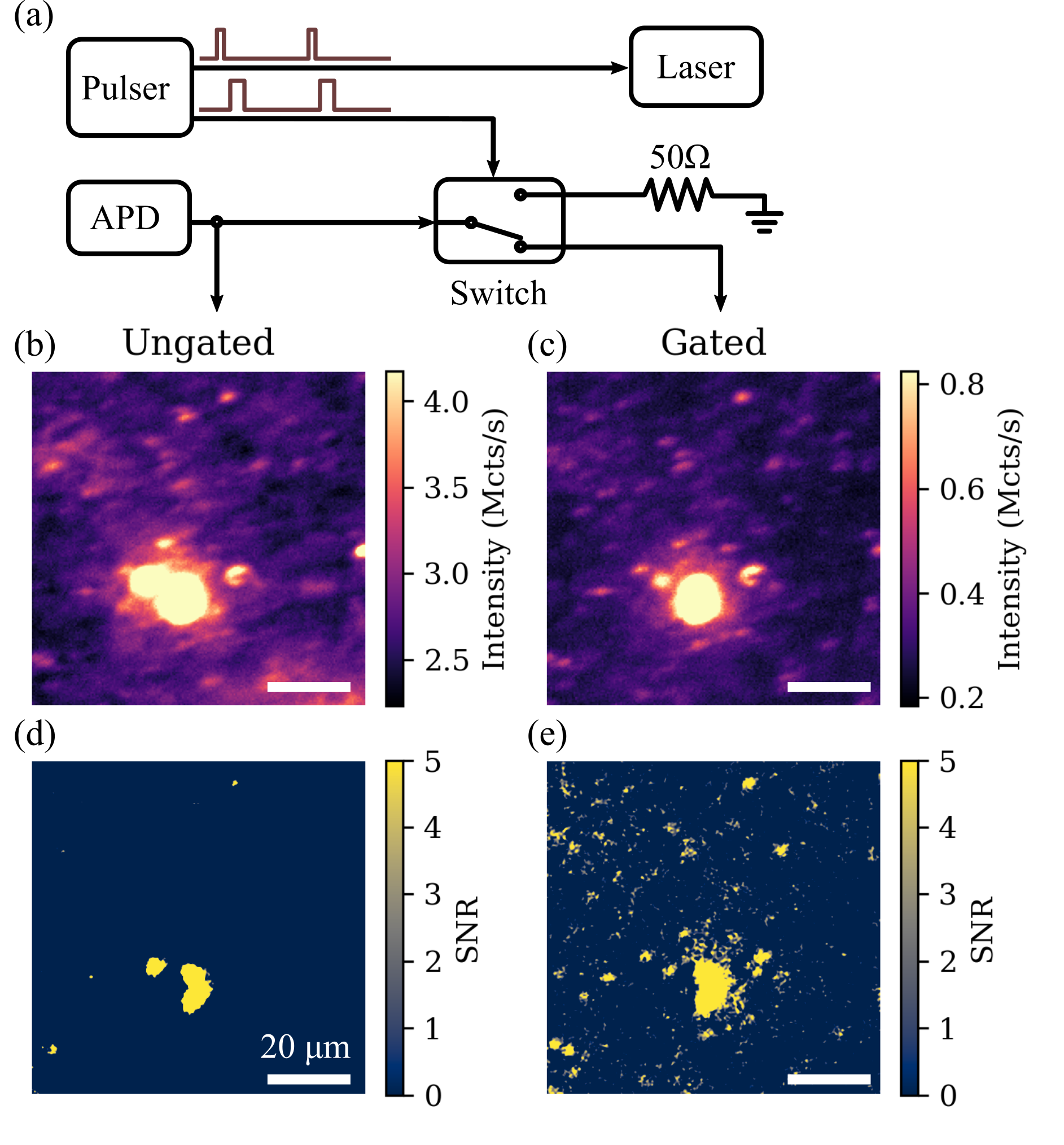}
\caption{
(a) Schematic of experimental setup for real-time gating without TCSPC. A pulse generator (\textit{Pulser}) is generating the trigger for the pulsed \textit{Laser}, as well as a related pulse of defined length and phase to trigger a fast \textit{Switch}. The switch routes the TTL signals from the \textit{APD} either into a $50\Omega$ termination or to the photon counting electronics. This method shows a similar real-time SNR enhancement as obtained via time-gating using TCSPC and post-processing.  
(b,c) Ungated and gated confocal scans of a $80\,$\textmu m$\times80\,$\textmu m region on the NC strip covered with FNDs, respectively.
(d,e) The resulting SNR calculated from two subsequent confocal scans, one with the MW frequency set on resonance and one with the MW off.
}
\label{fig:fig5}
\end{figure}

As applications like IVD are expected to be performed in a rapid fashion, which means in-situ evaluation, and with possibly less expensive hardware, we further demonstrate that time-gating can be performed in hardware in real-time and without the need for dedicated fast counting hardware. For this purpose, we record confocal maps of the FND covered NC with a relatively cheap objective lens with lower magnification (\textit{Olympus}, PLN 10X), where the purpose is the rapid detection and verification of the presence of FNDs. 
To accomplish real-time sensitivity enhancement, a fast switch (\textit{Mini-Circuits}, ZASWA-2-50DR+) is introduced to  allow time-gating of the single photon detector's TTL pulses, as shown in \hyperref[fig:fig5]{\autoref{fig:fig5}(a)}. 

One output of a digital pulse generator triggers the pulsed laser, and another output is connected to the fast switch, which routes the TTL pulses from the single-photon detector either to the digital counter card, or sends them into a $50\Omega$ termination. By correctly timing the trigger pulse for the laser in combination with the trigger pulse for the switch, we can achieve a similar result as by TCSPC and post-processing of the data. In \hyperref[fig:fig5]{\autoref{fig:fig5}(b-c)}, we show the resulting fluorescence maps of a NC strip covered with FNDs, without and with hardware time-gating, respectively. The two confocal maps show the same measurement, the only difference is, that one digital counter received all TTL pulses, and the other counter received only the time-gated TTL pulses. The pixel spacing is $0.4\,$\textmu m, and the integration time per pixel is $10\,$ms. 
In the ungated fluorescence map the background from the NC can clearly be seen to show stronger fluorescence, and overlap with the FNDs, whereas the time-gated fluorescence map shows a higher contrast from the (presumably) FNDs, and less background fluorescence by the NC. 
It is also interesting to see, that some of the fluorescent spots which appear in the ungated fluorescence map seem to vanish in the gated map, thus these spots are presumably also from background and do not relate to FNDs.

In order to certainly verify the presence of FNDs on the NC, the spin-dependent fluorescence variation of the NV center can be utilized. 
For this, the sample region is scanned twice, once with the MW off, and once with the MW set to the NV spin resonance ($2.87\,$GHz). From the resulting confocal maps, the spin-detection SNR at every point can be calculated using \autoref{eq:SNR}. 
In \hyperref[fig:fig5]{\autoref{fig:fig5}(d-e)} the resulting ungated and gated spin-detection, SNR maps can be found. To reduce small pixel-to-pixel variation noise from the two successive scans, a bicubic interpolation is applied to the resulting SNR maps. 
The SNR maps show, that the two big bright fluorescence spots in the center of the confocal scans, which are aggregates of FNDs, can be well distinguished with high SNR in both the ungated and gated cases. 
However, with the gated ODMR protocol, many of the fluorescent-weak FNDs could be revealed from the strong background environment through the temporal filtering, and promising for enhanced sensitivity in IVD applications.

Moreover, a cost-effective hardware time-domain filter (HTDF) was designed and produced to achieve high-resolution temporal gating of the NV fluorescence photons as to replace the expensive pulser and the TCSPC in future experiments. The HTDF circuit (see \hyperref[sec:appendix_ND]{Appendix \ref{sec:appendix_HTDF}}) is a fast switch for routing the photon signal from a photo detector to a counting device or to ground. The gate can be triggered \textit{on} by the trigger output of the pulsed laser. The \textit{on} and \textit{off} time can be changed by tuning a pair of adjustable resistors. With the HTDF, we are able to perform gated detection without the high-resolution pulser and the TCSPC, and it even allows for using a photodiode instead of the APD.


\section{Conclusion and outlook}

In conclusion, this study highlights the effectiveness of time-gating in enhancing the SNR and magnetic field sensitivity using CW-ODMR in NV centers in diamond, especially in high-background environments such as encountered in biological sensing and spin-enhanced IVD. The principle is first demonstrated in bulk diamond, where the main background contribution is by SiV centers, where the SNR and magnetic field sensitivity could be increased by a factor of 2, leading to a measurement speedup by a factor of 4. Then, in order to demonstrate the practical application in IVD, we show a strong SNR enhancement of detecting FNDs on a NC substrate, which is the typical scenario in IVD. Here we could demonstrate a significant improvement in SNR up to a factor of 4, and a resulting measurement time reduction by a factor of 16. Through careful optimization of time-gating parameters and laser repetition rates, we could show how to effectively mitigate background fluorescence, enhancing the detection capability of FNDs embedded in complex environments.
The method outlined is simply intensity based and does not rely on fitting or recording the actual TCSPC curve, hence we could also demonstrate a simple real-time implementation in hardware without the requirement for any dedicated devices. 
Additionally, e.g., in fragile biological environments, an additional advantage of pulsed excitation is the circumventing of potential sample heating associated with CW excitation.

The method and results here are not only limited to high-resolution confocal microscopy. The gating methods can also be applied to larger-area imaging and sensing using photodiodes for high-density NV samples,  as well as wide-field imaging and sensing with NV centers. In both scenarios, due to lower spatial filtering capabilities, the background contribution can be even more detrimental on the NV spin-detection. 
Furthermore,  we only used a binary encoding of the registered photons in our implementation so far, where the registered fluorescence was either used as-is, or discarded. A further strong improvement could be achieved by weighting each time bin to maximize the extracted information, e.g., using machine learning approaches \cite{Qian2021}.

In summary, while this study demonstrates the immediate benefits of time-gating for spin-enhanced biosensing, ongoing research and technological innovations hold promise for further enhancing the versatility and efficacy of NV center-based quantum sensing in diverse biomedical and environmental applications.

\section{Acknowledgments}
This study is financially supported by Beijing Natural Science Foundation (Z240006),the Smart Grid-National Science and Technology Major Project [2024ZD0803300] and National Natural Science Foundation of China (52172056).

\appendix

\section{Characterization of the nanodiamond sample}
\label{sec:appendix_ND}

The nanodiamond samples were prepared from milling type Ib high temperature high pressure bulk diamonds, and showed a typical diameter of $200\,$nm. NV centers in the FNDs were created by electron beam irradiation with kinetic energy of $12\,$MeV and subsequent $10\,$hours of annealing under $800^\circ$C in vacuum. 
\begin{figure}
  \centering
  \includegraphics[width=1\columnwidth]{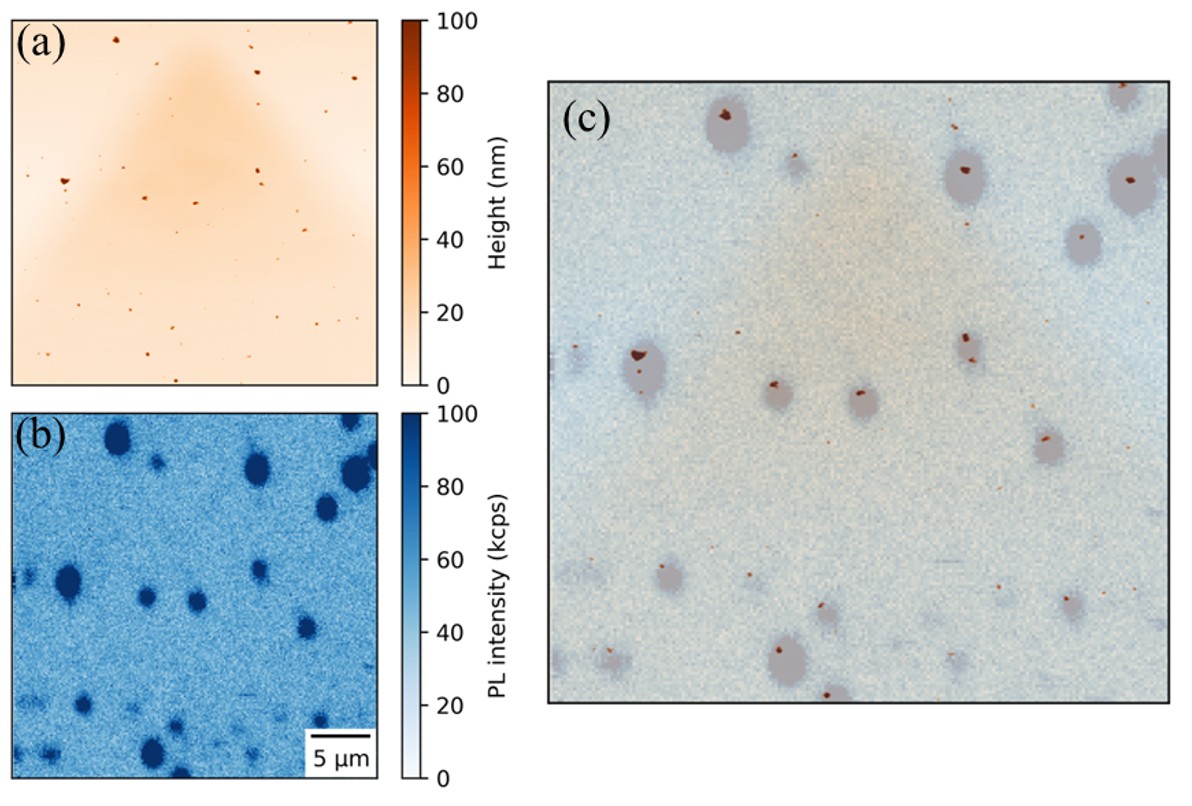}
\caption{(a) AFM scan of the nanodiamond sample. (b) Confocal scan of the same area as in (a), showing the fluorescence from the NV centers. (c) The overlap of the AFM and confocal scans, showing that most of nanoparticles were fluorescent with the typical NV fluorescence spectrum shown in \hyperref[fig:fig4]{\autoref{fig:fig4}(a)}. For resolving the FNDs, the AFM scan was set with $2048\times 2048$ pixels, which took a much longer time than the confocal scan ($200\times 200$ pixels). And with a slow sample drifting along the vertical axis, the accumulated displacement between the two scans can be observed in their overlap.}
\label{fig:fig_apdx_ND_afm}
\end{figure}

The produced NV nanodiamond sample was dissolved in de-ionized water. We used a pipette to deposit a few $\mu L$ of FND solution on a thin coverglass with spin coating and let it dry in air for a few minutes. The prepared FND sample was then characterized by a combined setup consisting of an atomic force microscope (AFM) and a confocal microscope\cite{sakar2021quantum, rogers2019single}. The AFM scan of the sample is shown in \hyperref[fig:fig_apdx_ND_afm]{\autoref{fig:fig_apdx_ND_afm}(a)}, where the typical size of the nanodiamonds is about $200\,$nm. The big particles are nanodiamond clusters formed during drying as water tension could pull some of FNDs together. The confocal scan of the same area is shown in \hyperref[fig:fig_apdx_ND_afm]{\autoref{fig:fig_apdx_ND_afm}(b)}, where the fluorescence from the NV centers can be seen. The overlap of the AFM and confocal scans is shown in \hyperref[fig:fig_apdx_ND_afm]{\autoref{fig:fig_apdx_ND_afm}(c)}, showing that most of nanoparticles were fluorescent and they showed the typical NV fluorescence spectra as the one in \hyperref[fig:fig4]{\autoref{fig:fig4}(a)}.

\section{Hardware Time Domain Filter}
\label{sec:appendix_HTDF}

\begin{figure*}[htbp]
    \centering
    \includegraphics[width=0.95\linewidth]{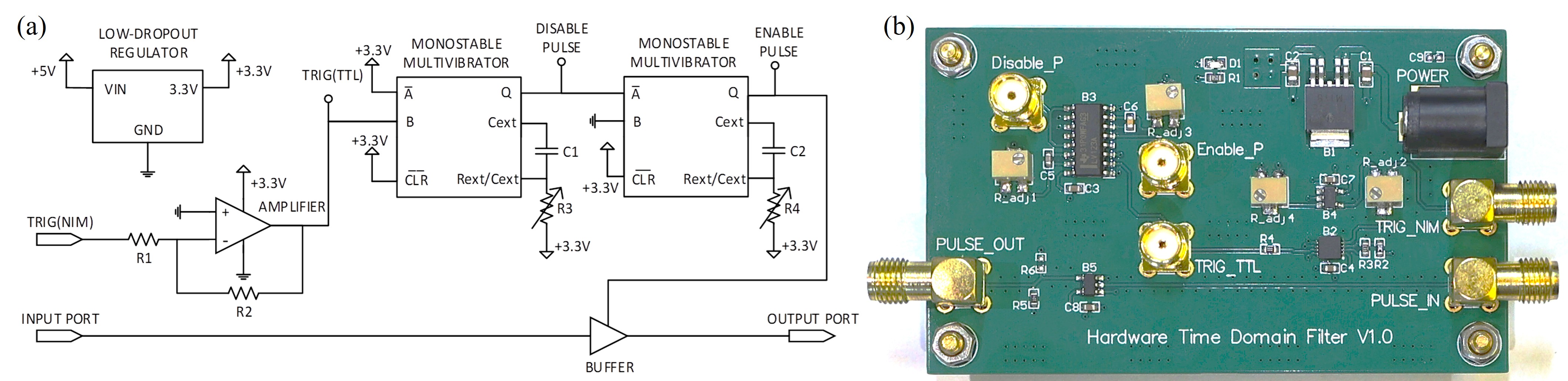}
    \caption{(a)Schematic circuit design of the Hardware Time Domain Filter. The “INPUT PORT” is used to receive electronic signal from a photo detector, and the “OUTPUT PORT” is used to send gated signal to a counting device. The gate-on can be triggered by either a NIM signal (at the “TRIG(NIM)” port) or a TTL signal (at the “TRIG(TTL)” port). The gate on and off times can be controlled by tuning resistor R3 and R4 respectively. (b)Printed circuit board layout of the HTDF($80mm\times48mm$)}
    \label{fig:fig_apdx}
\end{figure*}

The Hardware Time Domain Filter (HTDF) circuit was designed (schematic diagram shown in \hyperref[fig:fig_apdx]{\autoref{fig:fig_apdx}(a)}) and produced (actual photo shown in \hyperref[fig:fig_apdx]{\autoref{fig:fig_apdx}(b)}) to achieve high-resolution temporal gating of the NV fluorescence photons.
The HTDF uses two monstable multivibrators to create a fast switch. The first one is used to generate a trigger pulse with a tunable length ($\tau_c$) by tuning an adjustable resistor(R3 in \hyperref[fig:fig_apdx]{\autoref{fig:fig_apdx}(a)}), which is then used to trigger the other one for the genaration of gate "on" signal. The gate "on" time window ($1/f_L-\tau_c$) can be tuned by a second adjustable resistor (R4 in \hyperref[fig:fig_apdx]{\autoref{fig:fig_apdx}(a)}). The gate "on" signal is then used to control the switch ("BUFFER" in \hyperref[fig:fig_apdx]{\autoref{fig:fig_apdx}(a)}) , which routes the incoming TTL pulses from the APD either to floating or to the counting device.

The circuit operates with a 5 V power supply, regulated to 3.3 V via a low dropout (LDO) regulator. The trigger source (from a pulsed laser) can be a NIM signal or a TTL signal. As shown in \hyperref[fig:fig_apdx]{\autoref{fig:fig_apdx}(b)}, there are six SMA connectors in the printed circuit. The “PULSE\_IN” connects to a photo detector, and the “PULSE\_OUT” connects to a photo counting device. One can select either “TRIG\_NIM” or “TRIG\_TTL” to receive a laser pulse trigger. The “Disable\_P” and “Enable\_P” ports are used to monitor gate start ($\tau_c$) and gate on duration ($1/f_L-\tau_c$).

In brief, the circuit is a fast switch for routing the photon signal from a photo detector to a counting device or to ground. The gate can be triggered “on” by the laser pulse. The “on” and “off” time can be changed by tuning a pair of adjustable resistors (R3 and R4 in \hyperref[fig:fig_apdx]{\autoref{fig:fig_apdx}(a)}, R\_adj1 and R\_adj4 in \hyperref[fig:fig_apdx]{\autoref{fig:fig_apdx}(b)} respectively). With the HTDF, we are able to perform gated detection without the high-resolution pulser and the TCSPC, and it even allows for using a photodiode instead of the APD.

\section{Optimizing Repetition Rate}\label{app:rep_rate}
\begin{figure}[htbp]
  \centering
  \includegraphics[width=1\columnwidth]{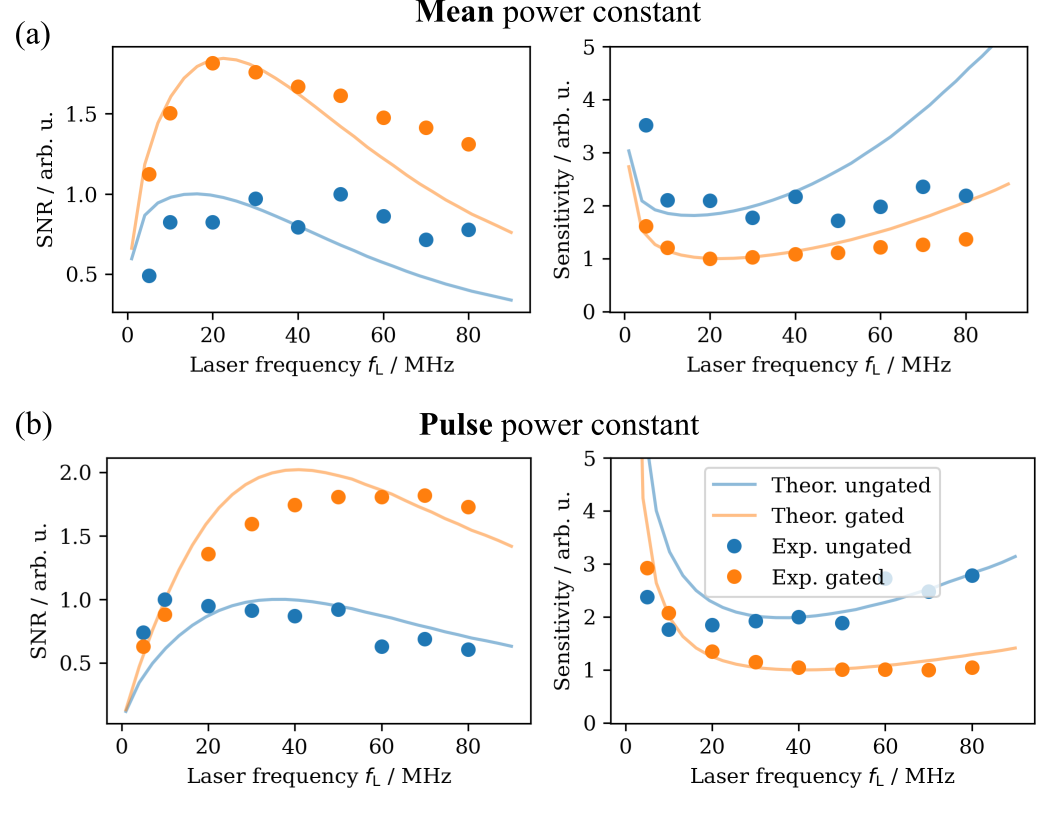}
\caption{
Laser repetition rate dependent SNR (left) and magnetic field sensitivity (right) without and with optimized time gating (data points) and theoretical approximation using a simple bi-exponential approximation (solid lines, see main text). The excitation laser power was set to be below saturation for a repetition rate of $40\,$MHz ($p_\text{L}\approx 150\mu\text{W}$), and in (a) the mean power of the pulsed excitation laser was kept constant, whilst in (b) the power per individual excitation pulse was kept constant.
}
\label{fig:fig_rep_rate_opt}
\end{figure}

For certain short-lived background sources, a good filtering of short background fluorescence, and a good SNR improvement can be achieved by lower laser repetition rates ($1/f_{\text{L}} \gg \tau_0$), however we also have to consider that higher laser repetition rates increases the total amount of collected photons in a certain measurement duration.

In the case of very low repetition rates $1/f_{\text{L}} \gg \tau_i$, meaning the NV center most likely decayed to its ground state can be re-excited by the next arriving laser pulse, there should be a linear relationship between $f_{\text{L}}$ and the collected photons $N_i$ (with $i\in[0,1]$) hence the SNR should scale with $\sqrt{f_{\text{L}}}$. 
However, at higher laser repetition rates ($1/f_{\text{L}} \gtrsim \tau_0$) the NV centers can show saturation, the time gating can be ineffective, and the re-excitation efficiency of the NV center can be decreased, meaning a larger contribution of the detected signal will stem from background noise. 
It is therefore essential to optimize the laser repetition frequency together with the time-gating window, in order to find the sweet spot of the repetition rate, and achieve maximal SNR enhancement.

In an actual sample, changing the laser repetition rate will have an influence on multiple physical processes, amongst others the NV ionization rate, the NV repolarization efficiency, as well as possible bleaching of background fluorescence. It is therefore important to experimentally verify the ungated and gated SNR for different laser repetition rates in a given measurement time, which can vary depending on the nature of the background, as well as NV centers. 

In order to properly compare the resulting SNR and sensitivity obtained by varying the repetition rate, we consider two cases: In the first case we keep the mean power constant, which can be crucial if the heating of the sample can be a problem, e.g., in the imaging of cells. 
On the other hand, we look at the case where the pulse power is set to be constant, e.g., if it is crucial to avoid damage to the sample from strong single laser pulses. 
The resulting SNR and sensitivity without and with time gating can be found by the data points in \autoref{fig:fig_rep_rate_opt}. The excitation laser power was set to be below saturation of the NV center, and the same mean laser power was chosen in both experiments for a repetition rate of $40\,$MHz ($p_\text{L}\approx 150\mu\text{W}$). Furthermore, the time gates have been optimized at each repetition frequency, as discussed before.

The solid lines in \autoref{fig:fig_rep_rate_opt} show the predicted results from a very simplified analytic model, which consists of the sum of two exponential decays with one fast exponential decay with time constant $\tau_\text{SiV}= 1.5\,$ns and one slower exponential decay with either $\tau_\text{NV0}= 12\,$ns, or $\tau_\text{NV1}= 11\,$ns, plus constant background contribution. The parameters were chosen from fitting bi-exponential decays to the observed TCSPC curves, and from these analytic decay curves the SNR and sensitivity are calculated. 
It should be emphasized, that this very simplified theoretical model does not take in account things such as saturation, bleaching or other more complex effects, and the curves are merely shown to represent the trend of the experimental data.

The results in \autoref{fig:fig_rep_rate_opt} show that the SNR and magnetic field sensitivity can have a strong dependence on the laser repetition rate, and thus it is not only important to find the right gating window  $\tau_\text{c}$, but also to optimize the laser repetition rate $f_{\text{L}}$ in order to maximize the NV spin-detection SNR and sensitivity.  
As we could demonstrate, the optimal gating window and repetition rate is dependent on a variety of factors, and needs to be deduced for each new environment setting. The optimal laser repetition frequency is especially dependent on the circumstances, and the resulting optimized excitation parameters depend on if either the mean laser power needs to be kept constant, or the pulse power should be kept constant.

\bibliography{bib.bib}
\bibliographystyle{apsrev4-2}

\end{document}